\documentclass[12pt]{article}
\usepackage{amsmath}
\usepackage{graphicx}
\begin{document}
\begin{center}
\section*{Hadronic Decays and Baryon Structure}
\end{center}
\begin{center}
{\bf T. Melde}, {\bf W. Plessas}, and {\bf B. Sengl}\\
\vspace{0.3cm}
Theoretical Physics, Institute for Physics, University of Graz,\\
Universit\"atsplatz 5, A-8010 Graz, Austria
\end{center}
\begin{abstract}
Relativistic constituent quark models generally describe three-quark
systems with particular interactions. The corresponding invariant mass eigenvalue
spectra and pertinent eigenstates should exhibit the multiplet structure 
anticipated for baryon resonances. 
Taking into account the flavour content, spin structure, and spatial distribution of the
baryon wave functions together with mass relations of the eigenvalues and decay 
properties of the eigenstates, we can link the theoretical mass eigenstates
with the experimentally measured resonances.
The resulting classification of baryon resonances 
differs in some respects from the one suggested by the Particle Data Group. 
With regard to the hadronic decay widths of light and strange baryon resonances
a consistent picture emerges only, if the classification includes two-star resonances.
\end{abstract}
\vspace{0.3cm}
\begin{center}
{\bf \large Introduction}
\end{center}
Constituent quark models (CQMs) for light and strange bary\-ons have seen a 
number of  important new developments over the last few years. 
Generally, CQMs are specified by a confining interaction and an interaction
responsible for the hyperfine splitting of the baryon spectra. There has been a number
of different implementations of hyperfine (and confining) interactions, and 
some prominent
models are based on one-gluon-exchange (OGE)~\cite{Capstick:1986bm},  
instanton-induced (II)~\cite{Loring:2001kx,Loring:2001ky}, and
Goldstone-boson-exchange (GBE) dynamics~\cite{Glozman:1998ag,Glozman:1998fs}. 

Recently, we have presented relativistic calculations of $\pi$ and $\eta$
decay widths of $N$ and $\Delta$ resonances within the point-form spectator model
(PFSM), and it has been seen that the experimental data are systematically 
underestimated~\cite{Melde:2005hy}.
Similar characteristics of these decay widths have been found
by the Bonn group following a completely different relativistic approach,
namely with the II CQM in the framework of the Bethe-Salpeter
equation~\cite{Metsch:2003ix,Metsch:2004qk}. 
Previous studies of mesonic baryon decays along CQMs essentially employed
nonrelativistic or relativised 
methods~\cite{Stancu:1989iu,Capstick:1994kb,Geiger:1994kr,Ackleh:1996yt,
Krassnigg:1999ky,Plessas:1999nb,Theussl:2000sj}. 
Our investigations of hadronic decays within the point form have now been extended
to the nonstrange decays of strange resonances~\cite{Sengl:2006Bled}. The
corresponding decay widths exhibit similar characteristics 
as the ones in the light sector~\cite{Melde:2005hy}. 
A specific interpretation has been reached with regard to the three
$\frac{1}{2}^-$ $\Sigma$ levels produced by CQMs (below 2 GeV): Only
the third excitation (in the GBE CQM) should be identified with the 
measured $\Sigma$(1750) resonance.
\vspace{0.2cm}
\begin{center}
{\bf \large Classification in Flavour Multiplets}
\end{center}

\renewcommand{\arraystretch}{1.2}
\begin{table}[t]
\begin{center}
\caption{
Suggested classification of experimentally seen baryons. The last column denotes the
multiplet number according to Guzey and Polyakov~\cite{Guzey:2005vz}. 
The superscripts denote the percentages of octet, singlet, and decuplet flavour
contributions in the respective states (specifically in case of the GBE CQM).
\label{tab:tab1}
}
\vspace{0.2cm}
{\begin{tabular}{@{}l llll c @{}}
\hline\hline
$(LS)J^P$& &&&&\#\\
\hline
Octets& &&&&\\
\hline
 $(0\frac{1}{2})\frac{1}{2}^+$
& $N(939)^{100}$  
& $\Lambda(1116)^{100}$
& $\Sigma(1193)^{100}$ &
$\Xi(1318)^{100}$
&1  \\
$(0\frac{1}{2})\frac{1}{2}^+$ 
& $N(1440)^{100}$  
& $\Lambda(1600)^{96}$
& $\Sigma(1660)^{100}$ 
& $\mbox{\boldmath$\Xi$}{\bf (1690)}^{100}$
&3\\
$(0\frac{1}{2})\frac{1}{2}^+$
& $N(1710)^{100}$  
&
& $\mbox{\boldmath$\Sigma$}{\bf (1880)}^{99}$ 
& &4 \\
$(1\frac{1}{2})\frac{1}{2}^-$
& $N(1535)^{100}$  
& $\Lambda(1670)^{72}$
& $\mbox{\boldmath$\Sigma$}{\bf (1560)}^{94}$
&  
&9  \\
$(1\frac{3}{2})\frac{1}{2}^-$
& $N(1650)^{100}$ 
& $\Lambda(1800)^{100}$ 
&$\mbox{\boldmath$\Sigma$}{\bf (1620)}^{100}$
& 
&14 \\
$(1\frac{1}{2})\frac{3}{2}^-$ 
& $N(1520)^{100}$
& $\Lambda(1690)^{72}$ 
& $\Sigma(1670)^{94}$ 
& $\Xi(1820)^{97}$
& 8 \\
$(1\frac{3}{2})\frac{3}{2}^-$
& $N(1700)^{100}$ 
&
& $\Sigma(1940)^{100}$ 
& &11 \\
$(1\frac{3}{2})\frac{5}{2}^-$ 
&$N(1675)^{100}$  
& $\Lambda(1830)^{100}$
& $\Sigma(1775)^{100}$ 
& $\mbox{\boldmath$\Xi$}{\bf (1950)}^{100}$
& 12  \\
\hline
Singlets& &&&&\\
\hline
$(0\frac{1}{2})\frac{1}{2}^+$ 
&&  $\mbox{\boldmath$\Lambda$}{\bf (1810)}^{92}$
&&&4 \\
$(1\frac{1}{2})\frac{1}{2}^-$
&& $\Lambda(1405)^{71}$ 
&&&6  \\
$(1\frac{1}{2})\frac{3}{2}^-$  
&& $\Lambda(1520)^{71}$
&&&7 \\
\hline
Decuplets& &&&&\\
\hline
$(0\frac{3}{2})\frac{3}2{}^+$
& $\Delta(1232)^{100}$ 
& $\Omega(1672)^{100}$
& $\Sigma(1385)^{100}$ 
& $\Xi(1530)^{100}$ 
& 2 \\
$(0\frac{3}{2})\frac{3}{2}^+$
& $\Delta(1600)^{100}$ 
&& $\mbox{\boldmath$\Sigma$}{\bf (1690)}^{99}$
&& 5  \\
$(1\frac{1}{2})\frac{1}{2}^-$
& $\Delta(1620)^{100}$  
&&  $\mbox{\boldmath$\Sigma$}{\bf (1750)}^{94}$
&& 10 \\
$(1\frac{1}{2})\frac{3}{2}^-$
& $\Delta(1700)^{100}$ 
& & & &13 \\
\hline\hline
\end{tabular}}
\end{center}
\end{table}

Motivated by the consistent picture that arose from the PFSM results
for hadronic decay widths we undertook a classification of the
mass-operator eigenstates into flavour
multiplets according to their most congruent behaviour of spatial densities,
spin as well as flavour content, mass relations, and decay properties. A
natural pattern of flavour multiplets emerges 
that comprises also experimentally less well
established (i.e., two-star) resonances. The resulting multiplets are
summarized in Table~\ref{tab:tab1}.  
In the first column the total spin and parity $J^P$ of the flavour multiplet
members are given as well as the total orbital angular momenta 
$L$ and total spins $S$ specifying their wave functions in the rest frame.
The bold-face entries denote states where our classification differs from
the one by the PDG~\cite{PDBook}, and the last column refers to the multiplet
number according to the classification of
Guzey and Polyakov~\cite{Guzey:2005vz}. This classification
is nearly identical to ours. The only exception is the $\Lambda(1810)$,
which turns out to be a flavour singlet (with a percentage of 92\%)
rather than a flavour octet.

The PDG suggests a classification of baryons without consideration
of one- and two-star resonances~\cite{PDBook}. The proposed scheme closely
resembles the one by Samios et al.~\cite{Samios:1974tw} postulated already in
1974, when many of the resonances known today have not yet been confirmed. In
the context of modern relativistic CQMs one learns that also less well established
resonances of two-star status should be included into a classification of
flavour multiplets. A prominent example is the $\Sigma$(1750), which is to be
identified only with the third $\frac{1}{2}^-$ excitation in CQMs and 
turns out to be in a flavour decuplet.

\begin{figure}
\begin{center}
\includegraphics[height=7.0cm]{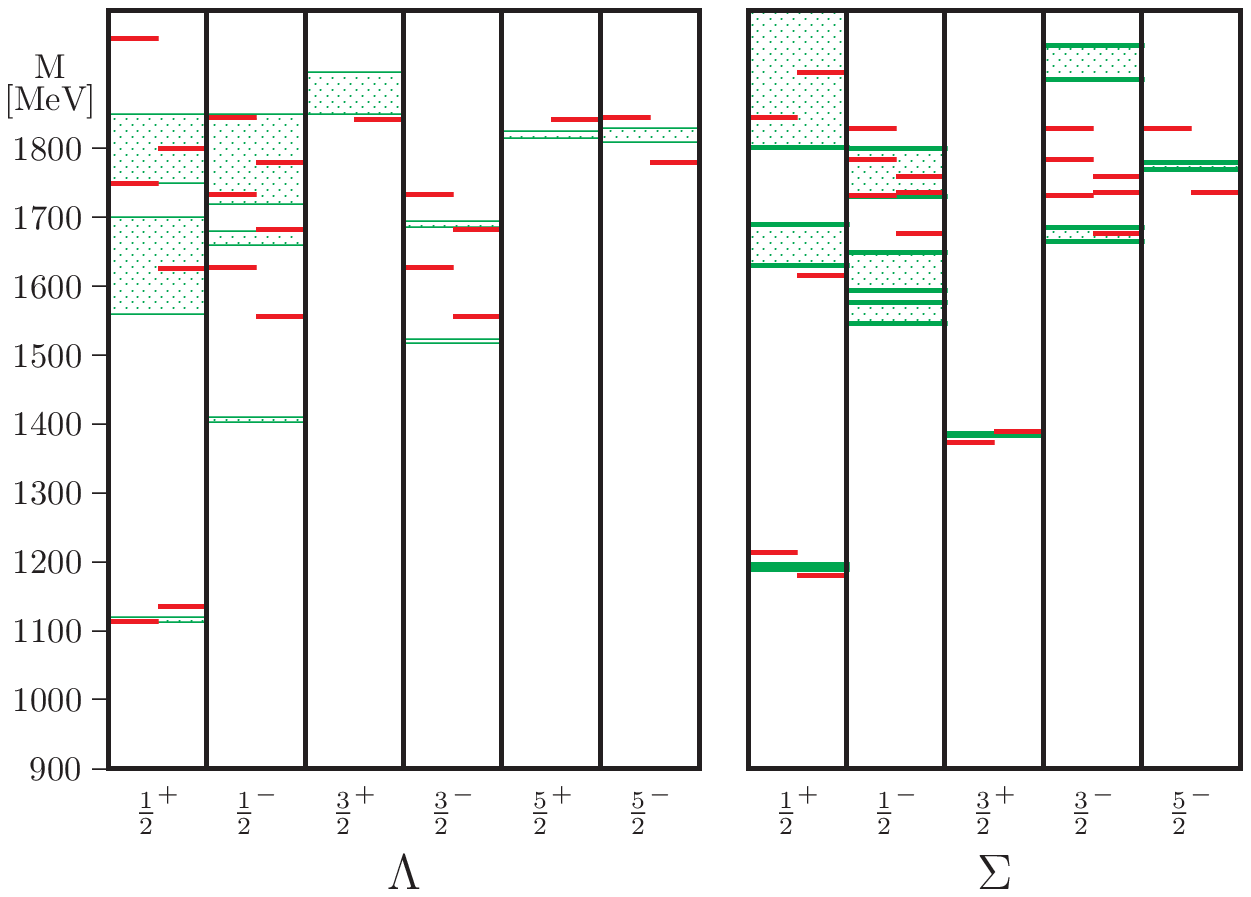}
\hspace{1mm}
\includegraphics[height=7.0cm]{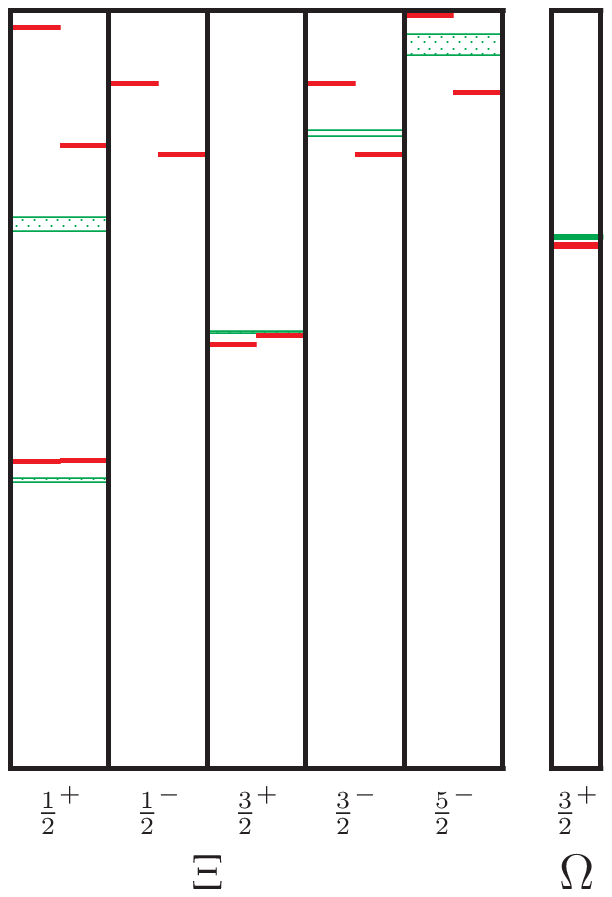}
\end{center}
\caption{$\Lambda$, $\Sigma$, $\Xi$, and $\Omega$ spectra for OGE 
(left) and GBE (right) CQMs. }
\label{fig:multiplet}
\end{figure}

The octet states in Table~\ref{tab:tab1} have a pure or very predominant octet
flavour content, with the notable exceptions 
of the $\Lambda(1670)$ and $\Lambda(1690)$; the latter couple strongly to
singlet states. The state $\Lambda(1810)$ is identified as a
(nearly pure) singlet state in concordance with a recent large-$N_c$ 
study~\cite{Matagne:2006zf}. The other two singlets exhibit considerable
admixtures of octet contributions, congruent with the singlet
contributions of their partners in the octet multiplets.
It should also be noted that the classification of the $\Xi$(1950) as 
$J^P=\frac{5}{2}^-$ is different from a recent one by
Zhou and Ma~\cite{Zhou:2006rr}, who classified it as a $J^P=\frac{3}{2}^-$.

In Fig.~\ref{fig:multiplet} we show all the experimental resonance states
(shadowed boxes) employed for the classification of the mass eigenstates
produced by the GBE and OGE CQMs (horizontal lines). Specifically in the
$\Sigma$ excitation spectrum with $J^P=\frac{1}{2}^-$ a natural explanation of
the states is found, if in addition to the $\Sigma$(1750) also the two
lower lying states are included, which are seen in experiment as two-star
resonances. It is interesting to note that the ordering of the three
$\Sigma$ states with $J^P=\frac{1}{2}^-$ is different in the two CQMs,
namely octet-decuplet-octet for the OGE and octet-octet-decuplet for the
GBE. Only in the $\Sigma$ spectrum with $J^P=\frac{3}{2}^-$ and the 
$\Xi$ spectrum with $J^P=\frac{1}{2}^-$we still observe more theoretical
states than experimentally seen resonances. However, at least in case of the
$\Sigma$ the PDG expects a resonance in the relevant mass range. 
\vspace{0.2cm}
\begin{center}
{\bf \large Summary}
\end{center}
We have investigated the properties of the light and strange baryons obtained
with the relativistic OGE and GBE CQMs. It has been found that
the CQMs provide a high degree of systematics with regard to the spectroscopy:
The invariant mass eigenstates yield a consistent pattern of flavour multiplets.
In particular, the $\Sigma$(1750) is identified as a flavour decuplet.
Additional two-star resonances can be interpreted consistently. A new
classification is reached differing in some respects from the one by the
PDG~\cite{PDBook}.

On the other hand, one faces difficulties in the description of baryon reactions.
In particular, the predictions of covariant decay widths along the PFSM cannot
explain all of the experimental results. Further relativistic studies are
necessary. In particular, investigations on the intricacies of the PFSM
construction~\cite{Melde:2004qu} might be of further relevance. For a more
refined approach the inclusion of explicit mesonic degrees of freedom appears
mandatory. Investigative coupled-channel calculations 
in a Poincar\'e-invariant quantum-mechanical framework
have already been performed in the meson 
sector~\cite{Krassnigg:2003gh,Krassnigg:2004sp}. However, the complexity of this
approach still prevents the application to baryons. For including 
mesonic degrees of freedom in the description of baryons, a promising first step
would be to take into account appropriate contributions (similar to the ones derived 
in Ref.~\cite{Canton:2000zf}) directly on the baryon-meson level. 

\vspace{0.3cm}
%
%\begin{center}
{\small This work was supported by the Austrian Science Fund (Projects 
FWF-P16945 and FWF-P19035).
B. S. acknowledges support through the Doktoratskolleg Graz
"Hadrons in Vacuum, Nuclei and Stars" (FWF-DK W1203). 
We like to thank F. Stancu
for pointing out the classification of the $\Lambda$(1810) as a singlet in the study of
Ref~\cite{Matagne:2006zf}.
}
%\end{center}
%
{\footnotesize
\renewcommand\refname{}

%\bibliographystyle{prsty}
%\bibliography{060721}

\vspace{-1.4cm}
\begin{thebibliography}{10}

\bibitem{Capstick:1986bm}
S. Capstick and N. Isgur, Phys. Rev. D {\bf 34},  2809  (1986).

\bibitem{Loring:2001kx}
U. Loering, B.~C. Metsch, and H.~R. Petry, Eur. Phys. J. A {\bf 10},  395
  (2001).

\bibitem{Loring:2001ky}
U. Loering, B.~C. Metsch, and H.~R. Petry, Eur. Phys. J. A {\bf 10},  447
  (2001).

\bibitem{Glozman:1998ag}
L.~Y. Glozman, W. Plessas, K. Varga, and R.~F. Wagenbrunn, Phys. Rev. D {\bf
  58},  094030  (1998).

\bibitem{Glozman:1998fs}
L.~Y. Glozman {\it et~al.}, Phys. Rev. C {\bf 57},  3406  (1998).

\bibitem{Melde:2005hy}
T. Melde, W. Plessas, and R.~F. Wagenbrunn, Phys. Rev. C {\bf 72},  015207
  (2005); Erratum, ibid. to appear  .

\bibitem{Metsch:2003ix}
B. Metsch, U. Loering, D. Merten, and H. Petry, Eur. Phys. J. A {\bf 18},  189
  (2003).

\bibitem{Metsch:2004qk}
B. Metsch, AIP Conf. Proc. {\bf 717},  646  (2004).

\bibitem{Stancu:1989iu}
F. Stancu and P. Stassart, Phys. Rev. D {\bf 39},  343  (1989).

\bibitem{Capstick:1994kb}
S. Capstick and W. Roberts, Phys. Rev. D {\bf 49},  4570  (1994).

\bibitem{Geiger:1994kr}
P. Geiger and E.~S. Swanson, Phys. Rev. D {\bf 50},  6855  (1994).

\bibitem{Ackleh:1996yt}
E.~S. Ackleh, T. Barnes, and E.~S. Swanson, Phys. Rev. D {\bf 54},  6811
  (1996).

\bibitem{Krassnigg:1999ky}
A. Krassnigg {\it et~al.}, Few Body Syst. Suppl. {\bf 10},  391  (1999).

\bibitem{Plessas:1999nb}
W. Plessas {\it et~al.}, Few Body Syst. Suppl. {\bf 11},  29  (1999).

\bibitem{Theussl:2000sj}
L. Theussl, R.~F. Wagenbrunn, B. Desplanques, and W. Plessas, Eur. Phys. J. A
  {\bf 12},  91  (2001).

\bibitem{Sengl:2006Bled}
B. Sengl, T. Melde, and W. Plessas, contribution to these proceedings

\bibitem{Guzey:2005vz}
V. Guzey and M.~V. Polyakov, hep-ph/0512355  (2005).

\bibitem{PDBook}
W.-M. {Yao} {\it et~al.}, {J. Phys. G} {\bf 33},  1  (2006).

\bibitem{Samios:1974tw}
N.~P. Samios, M. Goldberg, and B.~T. Meadows, Rev. Mod. Phys. {\bf 46},  49
  (1974).

\bibitem{Matagne:2006zf}
N. Matagne and F. Stancu, Phys. Rev. D {\bf 74},  034014  (2006).

\bibitem{Zhou:2006rr}
Q. Zhou and B.-Q. Ma, Eur. Phys. J. A {\bf 28},  345  (2006).

\bibitem{Melde:2004qu}
T. Melde, L. Canton, W. Plessas, and R.~F. Wagenbrunn, Eur. Phys. J. A {\bf
  25},  97  (2005).

\bibitem{Krassnigg:2003gh}
A. Krassnigg, W. Schweiger, and W.~H. Klink, Phys. Rev. C {\bf 67},  064003
  (2003).

\bibitem{Krassnigg:2004sp}
A. Krassnigg, Phys. Rev. C {\bf 72},  028201  (2005).

\bibitem{Canton:2000zf}
L. Canton, T. Melde, and J.~P. Svenne, Phys. Rev. C {\bf 63},  034004  (2001).

\end{thebibliography}
}

\end{document}